\documentclass[reprint,superscriptaddress]{revtex4-1}

\newcommand{\gv}[1]{\textcolor{black}{#1}}

\usepackage{graphicx}
\usepackage{amsmath}
\usepackage{amssymb}
\usepackage{color}
\usepackage{epstopdf}

\begin{document}
\title{Engineering sensorial delay to control phototaxis and emergent collective behaviors}

\author{Mite Mijalkov}
\affiliation{Soft Matter Lab, Department of Physics, Bilkent University, Cankaya, 06800 Ankara, Turkey}
\affiliation{UNAM -- National Nanotechnology Research Center, Bilkent University, Ankara 06800, Turkey}
\author{Austin McDaniel}
\author{Jan Wehr}
\affiliation{Department of Mathematics and Program in Applied Mathematics, University of Arizona, Tucson, Arizona 85721 USA}
\author{Giovanni Volpe}
\affiliation{Soft Matter Lab, Department of Physics, Bilkent University, Cankaya, 06800 Ankara, Turkey}
\affiliation{UNAM -- National Nanotechnology Research Center, Bilkent University, Ankara 06800, Turkey}



\begin{abstract}
Collective motions emerging from the interaction of autonomous mobile individuals play a key role in many phenomena, from the growth of bacterial colonies to the coordination of robotic swarms. For these collective behaviors to take hold, the individuals must be able to emit, sense and react to signals. When dealing with simple organisms and robots, these signals are necessarily very elementary, e.g. a cell might signal its presence by releasing chemicals and a robot by shining light. An additional challenge arises because the motion of the individuals is often noisy, e.g. the orientation of cells can be altered by Brownian motion and that of robots by an uneven terrain. Therefore, the emphasis is on achieving complex and tunable behaviors from simple autonomous agents communicating with each other in robust ways. Here, we show that the delay between sensing and reacting to a signal can determine the individual and collective long-term behavior of autonomous agents whose motion is intrinsically noisy. We experimentally demonstrate that the collective behavior of a group of phototactic robots capable of emitting a radially decaying light field can be tuned from segregation to aggregation and clustering by controlling the delay with which they change their propulsion speed in response to the light intensity they measure. We track this transition to the underlying dynamics of this system, in particular, to the ratio between the robots' sensorial delay time and the characteristic time of the robots' random reorientation. Supported by numerics, we discuss how the same mechanism can be applied to control active agents, e.g. airborne drones, moving in a three-dimensional space. Given the simplicity of this mechanism, the engineering of sensorial delay provides a potentially powerful tool to engineer and dynamically tune the behavior of large ensembles of autonomous mobile agents; furthermore, this mechanism might be already at work within living organisms such as chemotactic cells.
\end{abstract}

\maketitle

\section{Introduction}\label{sec:intro}

The interaction between several simple autonomous agents can give rise to complex collective behaviors. This is observed at all scales, from the organization of bacterial colonies \cite{Shapiro1998ARM,Berg2004Book} and the foraging of ants and bees \cite{Viswanathan2011Book} to the assembly of schools of fish \cite{Parrish2002BB} and the collective motion of human crowds \cite{Moussaid2009PRSB}. Inspired by these natural systems, the same principles have been applied to engineer autonomous robots capable of performing tasks such as search-and-rescue in disaster zones, surveillance of hazardous areas and targeted object delivery in complex environments \cite{Bonabeau2000Nature,Halloy2007Science,Sahin2008SI,Brambilla2013SI,Rubenstein2014Science,Werfel2014Science}. 

Complex behaviors can emerge even if each agent follows very simple rules, senses only its immediate surroundings and directly interacts only with nearby agents, without having any knowledge of an overall plan \cite{Schweitzer2003Book,Vicsek2012PR}. For example, while performing their swim-and-tumble motion, chemotactic bacteria are able to climb a chemotactic gradient, e.g. in order to move towards regions rich in nutrients, by simply adjusting their tumbling rate depending on the chemical concentration they sense \cite{Macnab1972PNAS,Berg2004Book}. Furthermore, by releasing chemoattractant molecules into their surroundings, they are capable of generating a chemical gradient around themselves to which other cells can respond, e.g. in order to create bacterial colonies \cite{Shapiro1998ARM}. Similarly, simple mechanisms are at work in the organization of flocks of birds, schools of fish, and herds of mammals, whereby complex collective behaviors result from each animal reacting to signals sent by its neighbors. A similar approach has also been fruitfully explored in order to build artificial systems with robust behaviors arising from interactions between very simple constituent agents \cite{Vicsek1995PRL,Bonabeau2000Nature,Dorigo2004AR,Chepizhko2013PRL,Palacci2013Science,Rubenstein2014Science,Werfel2014Science}.  Complex behaviors emerging from agents obeying simple rules have the advantage of being extremely robust: for example, even if one or more agents are destroyed, the others can continue to work together to complete the task at hand; agents can also be removed or added mid-task without significantly affecting the final result.

Here, we experimentally and theoretically demonstrate that it is possible to engineer the individual and collective behavior of autonomous agents whose motion is intrinsically noisy by making use of the delay in their sensorial feedback cycle. That is, we show how the delay between the time when an agent senses a signal and the time when it reacts to it can be used as a new parameter for the engineering of large-scale organization of autonomous agents. This proposal is inspired by the motion of chemotactic cells, which are able to climb a chemical gradient by adjusting a different parameter, i.e. their tumbling rate, in response to the concentration of molecules in their surroundings. We demonstrate that the collective behavior of a group of phototactic robots, capable of emitting a radially decaying light field, can be tuned from segregation to aggregation and clustering by controlling the delay with which they adjust their propulsion speed to the light intensity. More precisely, we show that this transition occurs as the ratio between the robots' sensorial delay time and characteristic time of their random reorientation crosses a certain critical value.

\section{Single agent}

\begin{figure*}[t!]
\includegraphics*[width=\textwidth]{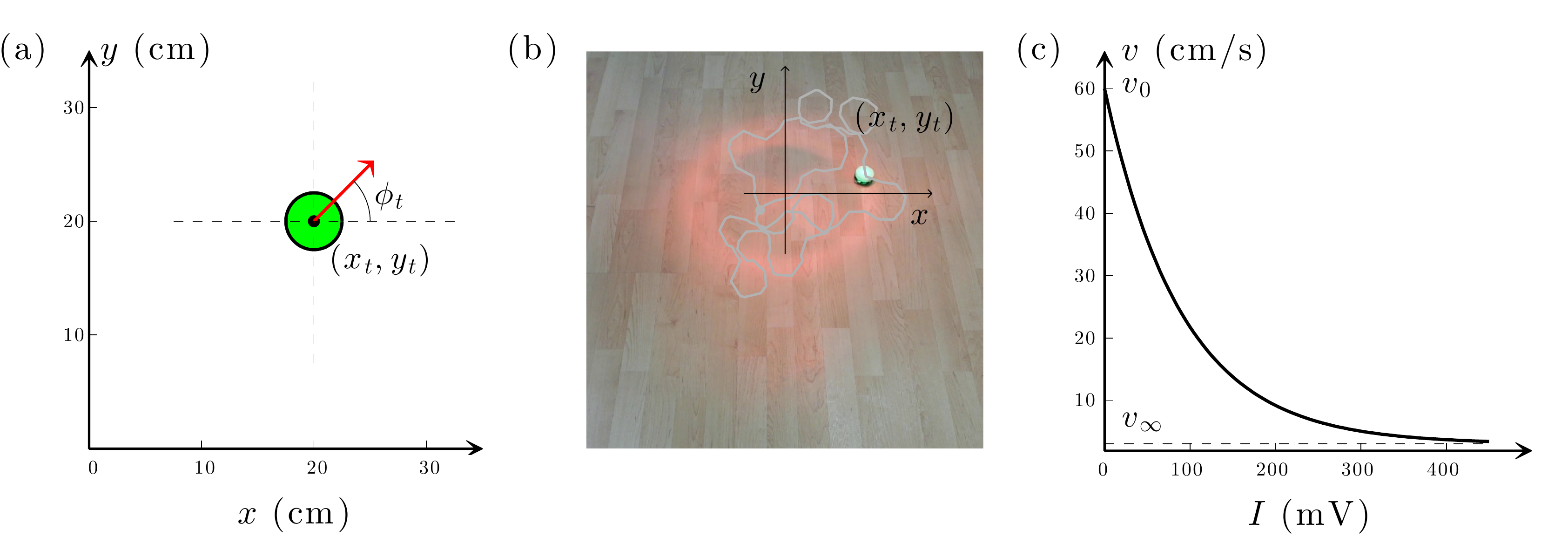}
\caption{
(a) An autonomous agent, whose position at time $t$ is $(x_t, y_t)$, moves with speed $v$ in the direction described by $\phi_t$, corresponding to its instantaneous orientation (arrow), which varies randomly with a characteristic time $\tau$. 
(b) Picture of a phototactic robot in a light gradient generated by an infrared lamp. The propulsion speed of the robot depends on the instantaneously measured light intensity, while its orientation changes randomly. A sample trajectory is shown by the gray solid line.
(c) Relation between the measured light intensity $I$ and the robot's speed $v$ [Eq.~(\ref{eq:v(I)})].}
\label{fig1}
\end{figure*}

We start by considering a single autonomous agent that moves in a plane and whose orientation is subject to noise. This happens naturally in the case of microswimmers ---~microscopic particles capable of self-propulsion such as motile bacteria and cells \cite{Ebbens2010SM,Volpe2014AJP}~--- as the direction of their motion changes randomly over time because of the presence of rotational Brownian motion \cite{Berg2004Book}. Similarly, autonomous robots, animals, and even humans can undergo a random reorientation when moving in the absence of external reference points (a striking example of this is an experiment where blindfolded people who were asked to walk in a straight line spontaneously moved along bent trajectories \cite{Souman2009CB}). Such motion is known as active Brownian motion and can be modelled by the following system of stochastic differential equations \gv{\cite{Schweitzer2003Book,howse2007self,peruani2007self,Volpe2014AJP}}:
\begin{equation}\label{eq:system:1}
\left\{
\begin{array}{ccl}
\displaystyle \frac{dx_t}{dt} 
&=&  
\displaystyle v \cos\phi_t \\[12pt]
\displaystyle \frac{dy_t}{dt} 
&=& 
\displaystyle v \sin\phi_t \\[12pt]
\displaystyle \frac{d\phi_t}{dt} 
&=&
\displaystyle \sqrt{\frac{2}{\tau}} \, \eta _t
\end{array}
\right.
\end{equation}
where $(x_t, y_t)$ is the position of the agent in the plane at time $t$, $\phi_t$ is its orientation, $v$ is its speed, $\tau$ is the reorientation characteristic time (i.e., the time after which the standard deviation of the agent's rotation is $1\,{\rm rad}$), and $\eta _t$ is a white noise driving the agent's reorientation, as shown in Fig.~\ref{fig1}a. \gv{The reorientation time $\tau$ can be associated to an effective reorientation diffusion constant $D_{\rm R}=\tau^{-1}$, which, in the case of microswimmers, often coincides with the rotational diffusion constant of the particle.}

Furthermore, we will assume that this agent moves in the presence of an external intensity field to which it reacts by adjusting its speed as a function of the instantaneous intensity it senses. We have realized this experimentally by using a phototactic robot (Elisa-3 \cite{elisa3}) moving within the light gradient generated by a $100\,{\rm W}$ infrared lamp, which emitted a radially symmetric light intensity radially decaying with a characteristic length $R = 35\,{\rm cm}$, as shown in Fig.~\ref{fig1}b. This robot measures the local light intensity $I_t =I(x_t, y_t)$ corresponding to its position $(x_t, y_t)$ at time $t$ using 8 infrared sensors evenly distributed around its circumference, and adjusts its propulsion speed $v\left( I \right)$ accordingly, while randomly changing its orientation with a characteristic reorientation time $\tau = 1\,{\rm s}$. Its motion can be described by modifying Eqs.~(\ref{eq:system:1}) as
\begin{equation}\label{eq:system:2}
\left\{
\begin{array}{ccl}
\displaystyle \frac{dx_t}{dt} 
&=&  
\displaystyle v\left( I_t \right) \cos\phi_t \\[12pt]
\displaystyle \frac{dy_t}{dt} 
&=& 
\displaystyle v\left( I_t \right) \sin\phi_t \\[12pt]
\displaystyle \frac{d\phi_t}{dt} 
&=&
\displaystyle \sqrt{\frac{2}{\tau}} \, \eta _t
\end{array}
\right.
\end{equation}
Fig.~\ref{fig1}b shows also a sample trajectory (line) superimposed onto the picture of the robot. The function $v(I)$ is plotted in Fig.~\ref{fig1}c; its functional form is
\begin{equation}\label{eq:v(I)}
v(I) = (v_{\rm 0}-v_{\infty}) e^{-I/I_{\rm c}} + v_{\infty} \; ,
\end{equation}
where $v_{\rm 0} = 60\,{\rm cm\, s^{-1}}$ is the maximum speed (corresponding to a null intensity), $I_{\rm c} = 90\,{\rm mV}$ is the characteristic intensity scale (measured in volts) over which the velocity decays, and $v_{\infty} = 3\,{\rm cm\, s^{-1}}$ is the residual velocity (in the limit of infinite light intensity). It can be seen in Fig.~\ref{fig1}b that the runs between consecutive turns are longer in the low-intensity (high-speed) regions while they are shorter in the high-intensity (low-speed) regions. The result is that over a long period of time, the robot spends more time in the high-intensity regions. As we will see, this behavior is in agreement with our theoretical results given in Eq.~(\ref{eq:density}). This is also in agreement with the behavior of chemotactic cells whose explorative behavior decreases when they reach regions with ideal conditions  and reduce their locomotion activity in favor of other metabolic activities \cite{Berg2004Book}.

\begin{figure}[h!]
\includegraphics[width=.5\textwidth]{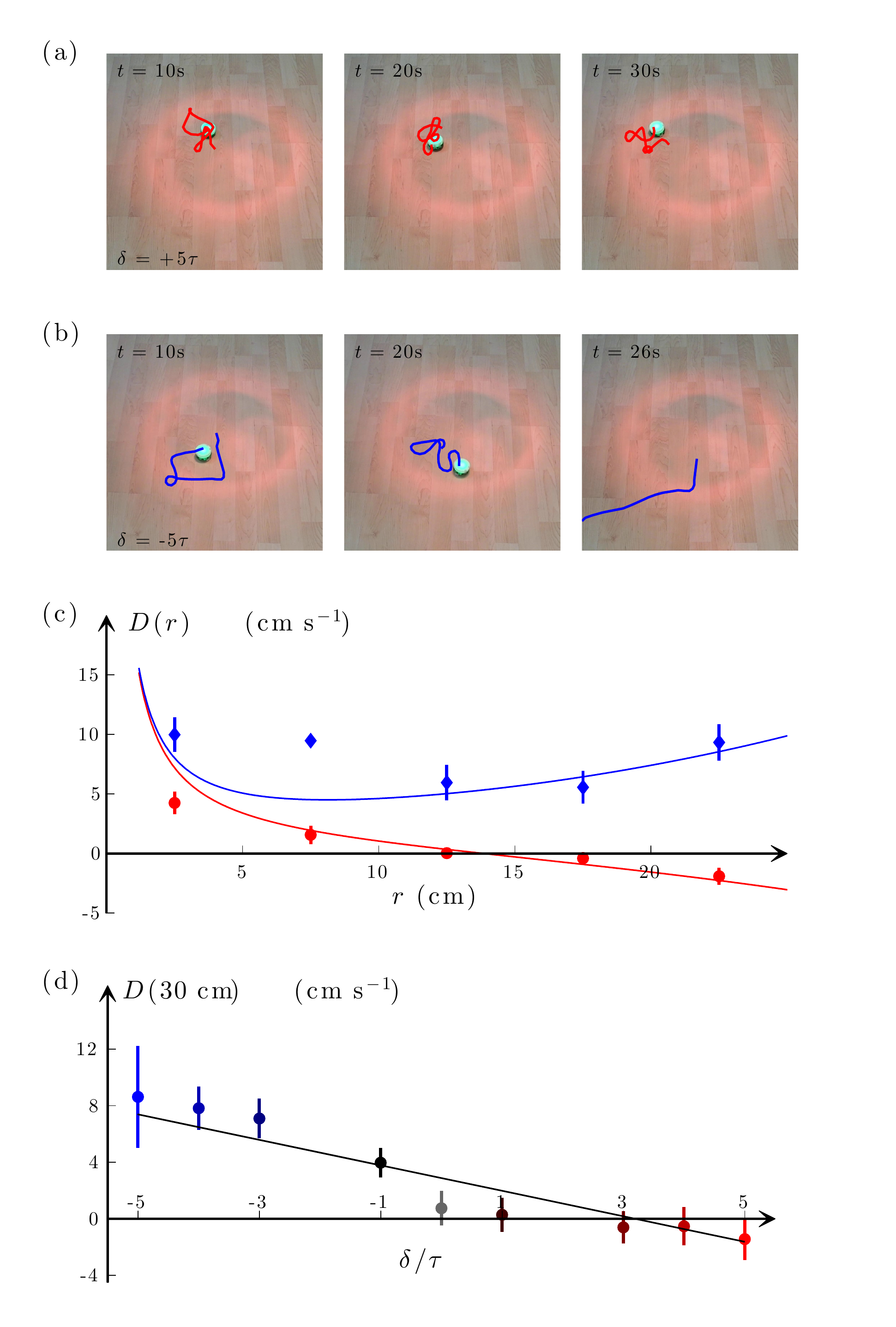}
\caption{The long-term behavior of a robot in the light gradient generated by an infrared lamp changes depending on the delay with which it adjusts its speed in response to the sensorial input, i.e. the measured light intensity. The sensorial delay was introduced by linearizing the measured light intensity as a function of time and by extrapolating its past/future value. (a) For positive delays ($\delta = +5 \tau$), the tendency of the robot to move towards the high-intensity (low-speed) regions is enhanced, when compared to the case without delay presented in Fig.~\ref{fig1}b.
(b) For negative delays ($\delta = -5 \tau$) the robot tends to move towards the low-intensity (high-speed) regions.
In both cases, the trajectories are shown for a period of $10\,{\rm s}$ preceding the time indicated on the plot and the robot is shown at the final position.
\gv{(c) Radial drift $D(r)$ calculated according to Eq.~(\ref{eq:Dr}) from a 40-minute trajectory for the cases of positive (circles) and negative (diamonds) delays. (d) Radial drift calculated according to Eq.~(\ref{eq:Dr}) when the robots are at $30\,{\rm cm}$ from the center of the illuminated area as a function of $\delta/\tau$. The solid lines in (c) and (d) correspond to the theoretically predicted radial drifts given by Eq.~(\ref{eq:drift}).}}
\label{fig2}
\end{figure}

We now proceed to add a delay $\delta$ in the agent's response to the measured intensity, which is the main novelty of our work. With this addition, the equations describing the motion of the robot become:
\begin{equation}\label{eq:system:delay}
\left\{
\begin{array}{ccl}
\displaystyle \frac{dx_t}{dt} 
&=&  
\displaystyle v\left( I_{t-\delta} \right) \cos\phi_t \\[12pt]
\displaystyle \frac{dy_t}{dt} 
&=& 
\displaystyle v\left( I_{t-\delta} \right) \sin\phi_t \\[12pt]
\displaystyle \frac{d\phi_t}{dt} 
&=&
\displaystyle \sqrt{\frac{2}{\tau}} \, \eta _t
\end{array}
\right.
\end{equation}
The idea of introducing a sensorial delay is inspired by the way in which bacteria react to a chemotactic gradient; in fact, chemotactic bacteria make a comparison of the number of molecules they detect around themselves at consecutive times in order to decide how to adapt their motion \cite{Segall1986PNAS,Macnab1972PNAS,Berg2004Book}. The presence of sensorial delays is typically ignored, or treated as a nuisance to be controlled \cite{Chen2011SCTS}, while only few theoretical works have considered its possible constructive effects but in situations different from the one studied in this work \cite{Forgoston2008PRE,Sun2014PRE}. By introducing a delay long enough so that the robot has enough time to randomize its direction of motion before responding to the sensorial input by changing its speed, we can observe that the motion becomes more directed towards the high-intensity (low-speed) region, as can be observed by comparing the trajectories in Fig.~\ref{fig2}a (with delay $\delta = +5\tau$) to that in Fig.~\ref{fig1}b (without delay).

Things become even more interesting if a ``negative" delay is introduced, i.e. if a prediction of the future measured intensity is employed to determine the current robot speed. \gv{While it is straightforward to see how a positive delay is introduced (e.g. by a delay in the transmission of the signal or by a lapse time before reacting to the signal), the introduction of a negative delay is less intuitive. In fact, a negative delay can be rationalized as a prediction of the future state of the system, which can be done based on the signal received up to the present time. For example, in the case of our robots, a negative delay is introduced by linearizing the light intensity measurement as a function of time and extrapolating it into the future, i.e. $I(t-\delta) \approx I(t) - \delta I'(t)$, where both $I(t)$ and $I'(t)$ are known at time $t$; higher order predictor algorithms are also possible making use of more information about the evolution of the intensity measured up to the present.} We show the corresponding trajectory in Fig.~\ref{fig2}b, where $\delta = -5\tau$. In this case, the robot escapes from the high-intensity region and moves towards the edge, where the infrared lamp intensity is lower (and the speed higher). 

In order to quantify these observations, we have measured the effective radial drift of the robots, which is calculated \cite{Pesce2013NC} as 
\begin{equation}\label{eq:Dr}
D(r) = \frac{1}{\Delta t} \langle r_{n+1} - r_{n} \mid r_{n} \cong r \rangle \; ,
\end{equation}
where $r$ is the radial coordinate, $r_n$ are samples of the robot's radial position and $\Delta t$ is the time step between samples. The results are shown in Figs.~\ref{fig2}c and \ref{fig2}d. For positive delay (red circles), the negative drift for large radial distance shows that the robot tends to move towards the central high-intensity region. For negative delay (blue diamonds), the positive drift shows that the robot escapes from the central high-intensity region. We have also theoretically calculated the radial drift for an autonomous agent whose motion is governed by Eqs.~(\ref{eq:system:delay}) (see Appendix~\ref{app:mathematicalderivation}), obtaining
\begin{equation}\label{eq:drift}
D(r) = \frac{\tau}{2} \left( 1 - \frac{\delta}{\tau} \right) v(r) \frac{d v}{d r} (r) + \frac{\tau v(r)^2}{ r} \; ,
\end{equation}
where $v(r) = v(I(r))$ and we have assumed a radially symmetric intensity distribution. The solid lines plotted in Figs.~\ref{fig2}c and \ref{fig2}d show that there is a good agreement between these theoretical predictions and the experimentally measured data. We have further corroborated these results with numerical simulations, whose results are shown in Fig.~S1 in the Supplementary Information and are in good agreement with the experimental results shown in Fig.~\ref{fig2}. The numerical simulations were performed by solving the finite difference approximation of  Eqs.~(\ref{eq:system:delay}) \cite{Kloeden1992Book,Volpe2014AJP}. The delayed sensorial measurement was evaluated by Taylor-expanding the measured intensity about the agent's location and extrapolating the corresponding past/future value.

\begin{figure}
\includegraphics[width=.5\textwidth]{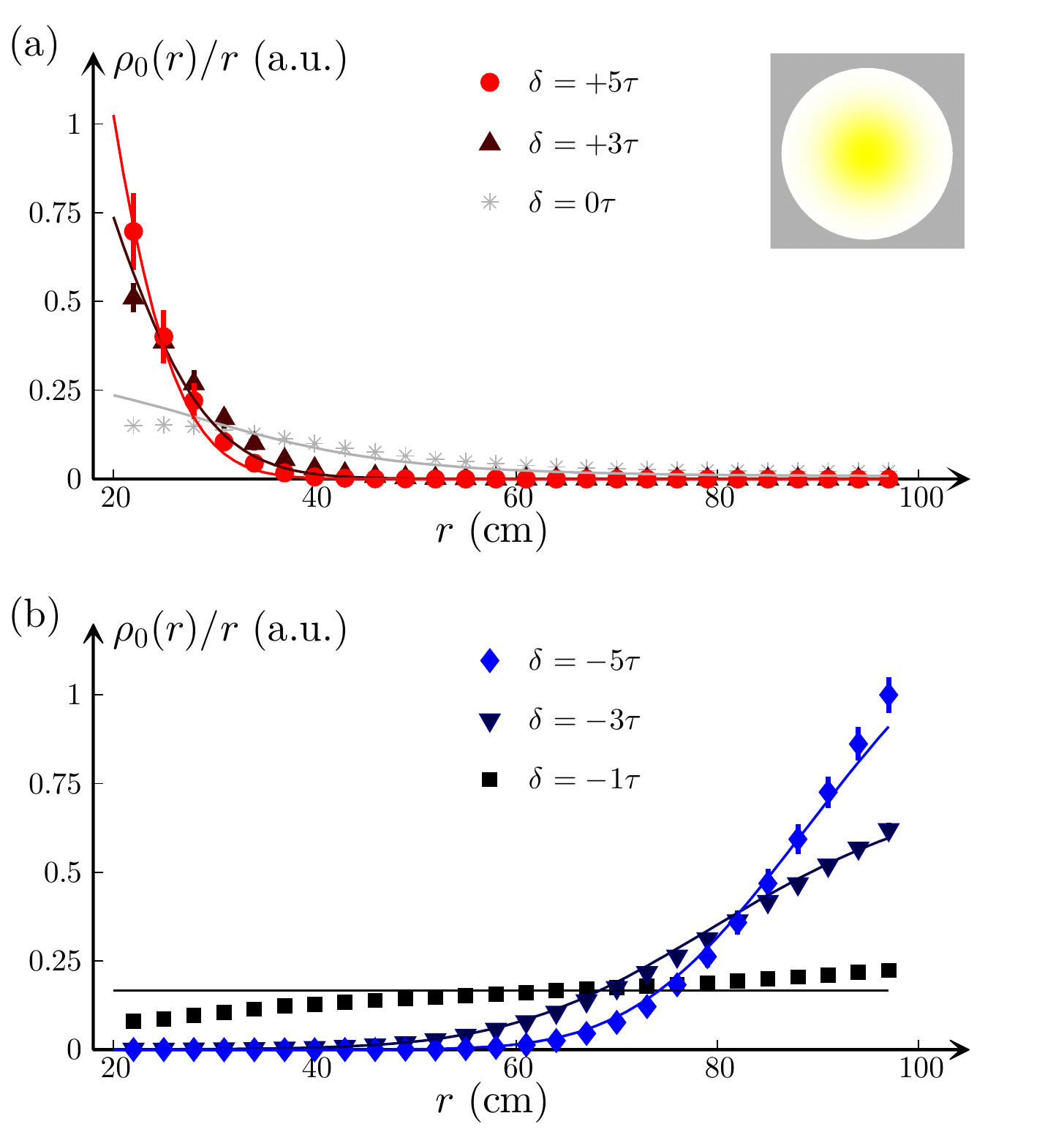}
\caption{Theoretically predicted radial probability distribution of the position of an agent [Eq.~(\ref{eq:density})] moving in a radial intensity field (inset in (a), the agent is confined in a circular well with radius $100\,{\rm cm}$ indicated by the gray border) as a function of the sensorial delay time: (a) for $\delta > -\tau$ the agent tends to spend more time in the low-speed (high-intensity) central region; (b) for $\delta < -\tau$ the agent spends more time in the high-speed (low-intensity) peripheral region; for $\delta = -\tau$ the probability distribution is uniform (black line). These results are corroborated by numerical simulations of autonomous agents shown by the symbols. For each case, we have simulated a very long trajectory ($10^8\,{\rm s}$) to obtain an accurate and smooth distribution.}
\label{fig3}
\end{figure}

We can also theoretically derive the approximate steady-state probability distribution of the agent's position (see Appendix~\ref{app:mathematicalderivation}), which exists and equals
\begin{equation}\label{eq:density}
\rho_0(x,y) = \frac{1}{N \; v(x,y)^{1+ \frac{\delta}{\tau}}} \; ,
\end{equation} 
provided that the normalization constant 
$$N = \int v(x,y)^{-\left(1+ \frac{\delta}{\tau}\right)}\,dx\,dy < \infty.$$ 
Eq. (6) confirms our initial observations that the larger the positive delay is (solid lines in Fig.~\ref{fig3}a), the more time the agent spends in the low-speed (high-intensity) regions.  On the other hand, the more negative the delay is (solid lines in Fig.~\ref{fig3}b), the more time the agent spends in the high-speed (low-intensity) regions. Interestingly, we note that there is a cutoff value at $\delta=-\tau$ for which the probability distribution of the agent is uniform (black solid line in Fig.~\ref{fig3}b). We have further corroborated these results with numerical simulations shown by the symbols in Figs.~\ref{fig3}a and \ref{fig3}b.

\gv{We emphasize that the qualitative change of the particle's behavior occurs at a negative delay, i.e. $\delta = -\tau$.  Introduction of negative delays is thus crucial for the described transition.  On the other hand, positive delays also influence the system's behavior strongly.  While without delay the particle spends more time in slow regions, a positive delay makes this tendency more pronounced, as seen clearly at the quantitative level from Eq.~(\ref{eq:density}).  This tendency persists, albeit in a weaker form, for small negative delays $-\tau < \delta < 0$ and gets reversed at the critical value $\delta = -\tau$.}

\section{Multiple agents}

\begin{figure}
\includegraphics[width=.5\textwidth]{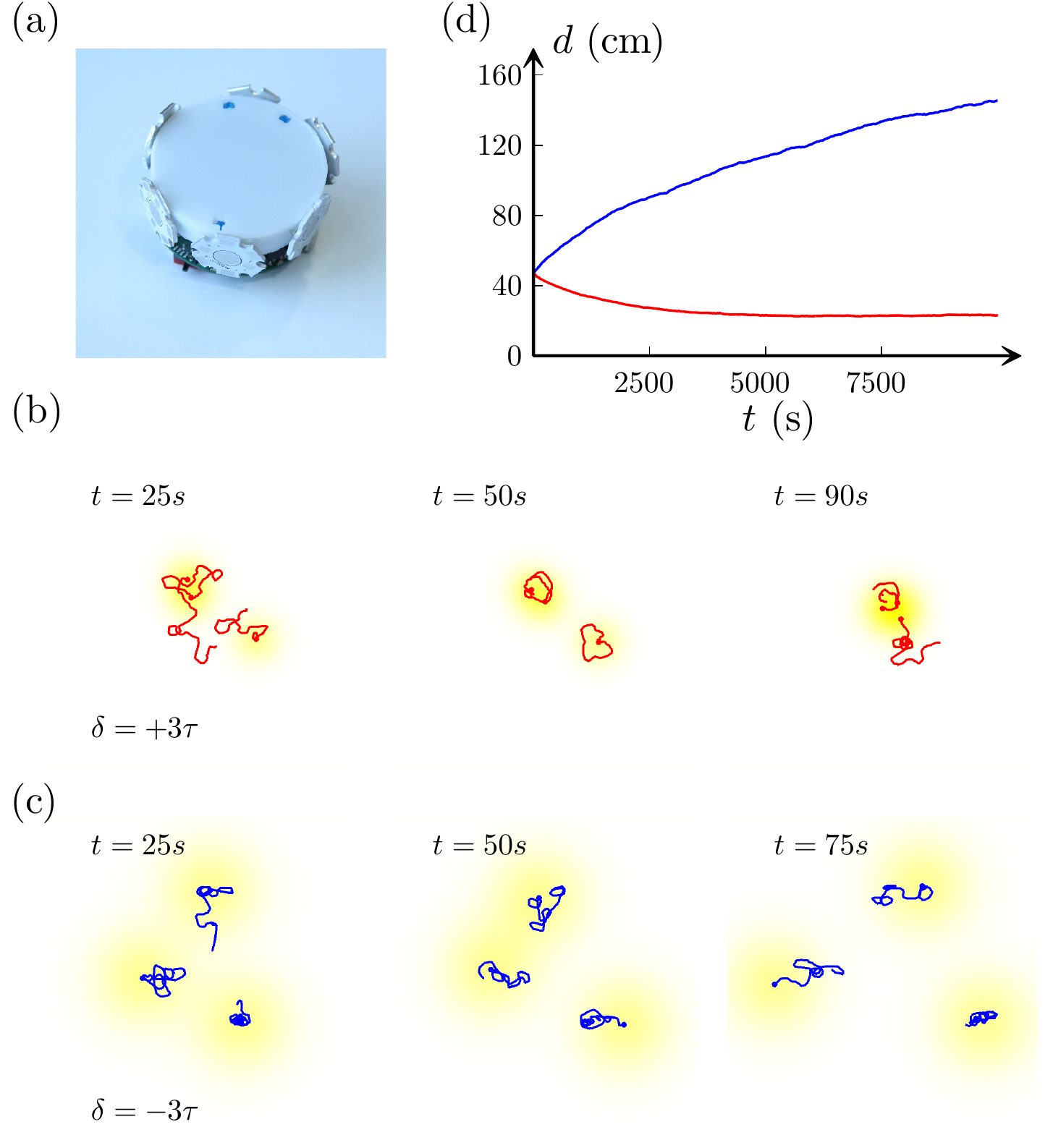}
\caption{(a) Picture of a phototactic robot equipped with six infrared LEDs so that it can emit a radially decaying light intensity around itself.
(b) A group of three such robots, which adjust their speed as a function of the sensed light intensity, aggregate and form a dynamic cluster if their sensorial delay is positive ($\delta = + 3 \tau$)
and (c) segregate if it is negative ($\delta = - 3 \tau$).
\gv{In each panel in (b) and (c) the trajectories are shown for a period of $10\,{\rm s}$ preceding the time indicated on the plot and the dot indicates the final position of the robot.}
(d) Average distance $d$ between agents in a group of three simulated autonomous agents as a function of time: for positive delays, as the agents tend to come together and form a cluster, their average distance decreases over time; for negative delays, as the agents move apart and explore a larger area, their average distance increases.}
\label{fig4}
\end{figure}

We can now build on these observations to engineer the large-scale organization of groups of robots. In order to do this, each robot must be able not only to sense the local intensity, but also to create a luminosity field. \gv{Thus, we have equipped each robot with 6 LEDs evenly placed around its circumference (EDEI-1LS3), as shown in Fig.~\ref{fig4}a, which emit infrared light (wavelength $850\,{\rm nm}$) so that each robot generates a decaying light intensity around itself.} \gv{The LEDs are arranged so that the robot measures only the light intensity emitted by the other robots.} A phototactic robot capable of measuring this light intensity will be able to move in the resulting field similarly to the case discussed above, i.e. that of the light intensity generates by a static infrared lamp. We stress that each robot only measures the local intensity without being aware of the positions of the other robots. 

\gv{We have experimentally studied how three autonomous robots organize by reacting to the cumulative light field created by all of them as a function of their sensorial delay.} For a positive sensorial delay ($\delta = +3 \tau$, Fig.~\ref{fig4}b), the three robots gradually move towards each other and form a dynamic cluster, which remains stable over time. A single robot's tendency to spend more time in the high-intensity regions when there is positive delay leads to multiple robots forming clusters because of their preference for high-intensity regions. For a negative delay ($\delta = -3 \tau$, Fig.~\ref{fig4}c), the three robots tend to move away from each other, dispersing and exploring a much larger area. In order to understand this behavior in a more quantitative way, we have also simulated \gv{a larger number of trajectories for a group of three agents} and plotted the average distance between the agents as a function of time for various sensorial delays. The results are reported in Fig.~\ref{fig4}d: for positive delays, as the agents tend to come together and form a cluster, their average distance decreases over time; for negative delays, as the agents move apart and explore a larger area, their average distance increases. \gv{The qualitative change of the agents' behavior occurs at a strictly negative value of the dimensionless parameter $\delta / \tau = - 1$ (see Eq.~(\ref{eq:density})). While introduction of negative delays is thus crucial for the described transition from aggregation to segregation, positive delays also influence the system's behavior strongly by enhancing the tendency of the agents to aggregate.} Importantly, not only a light field, but any radially decaying scalar (e.g. chemical, acoustic) field created by the autonomous agents can be used in order to achieve this kind of control over their behavior.

\begin{figure}
\includegraphics[width=.5\textwidth]{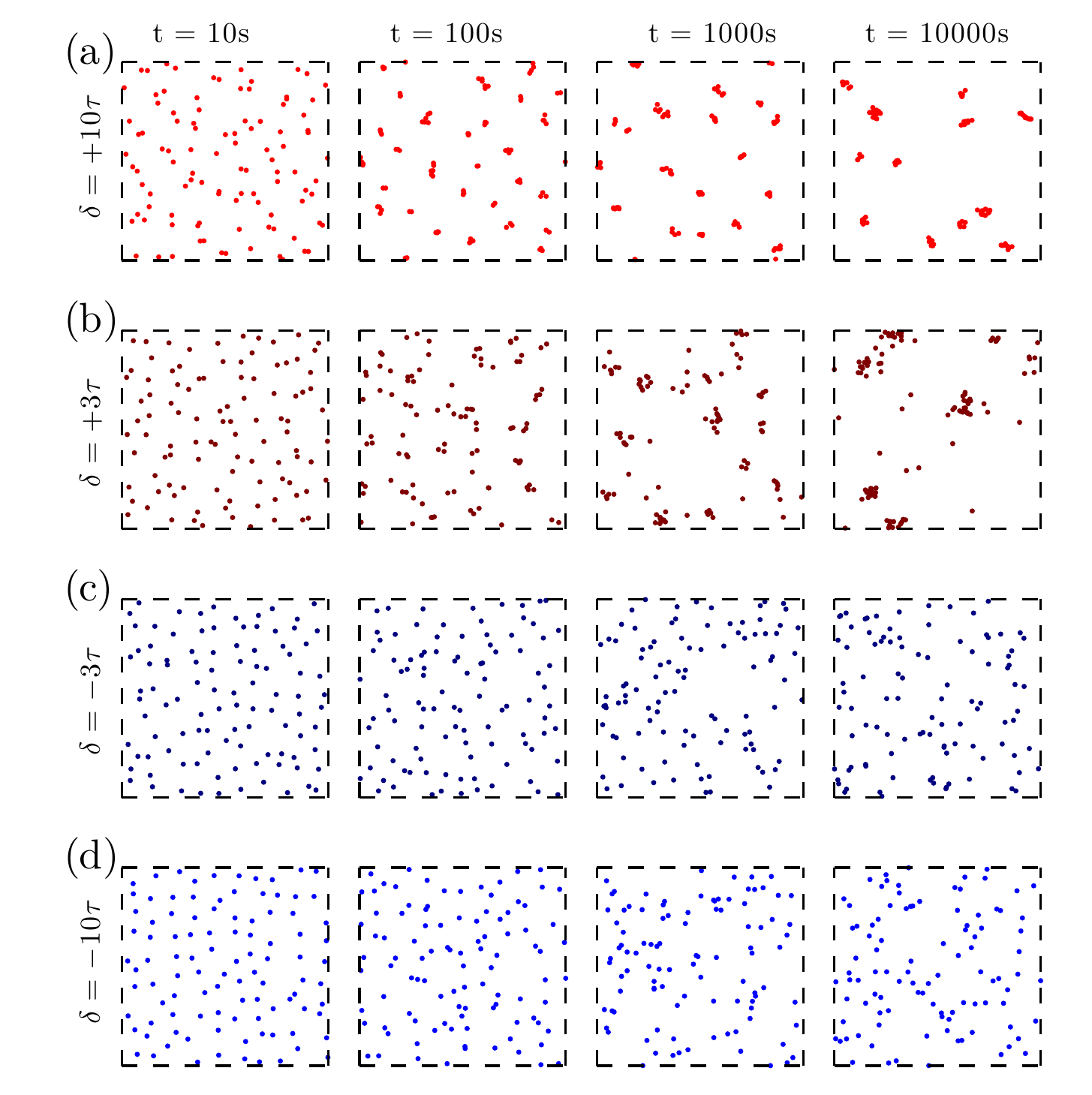}
\caption{Simulation of the long-term behavior of an ensemble of 100 autonomous agents that emit a radially decaying intensity field and adjust their speed depending on the measured local intensity. Depending on the sensorial delay, the long-term behavior and large-scale organization are significantly different. (a)-(b) In the case of positive delays, the agents come together and form metastable clusters. (c)-(d) In the case of negative delays, they explore the space, staying away from each other.}
\label{fig5}
\end{figure}

In order to explore the scalability of this mechanism, we have simulated the behavior of an ensemble of 100 robots. Each robot emits around itself a Gaussian intensity field that decays radially, and responds to the locally measured cumulative intensity by adjusting its speed. The long-term behavior and the large-scale organization of these ensembles of agents significantly depend on the sensorial delay, as shown in Fig.~\ref{fig5}. For positive delay, they move collectively by forming clusters (Figs.~\ref{fig5}a and \ref{fig5}b). On the other hand, for negative delays, they move away from each other in order to reduce the intensity each of them measures and are thus able to explore the space more effectively (Figs.~\ref{fig5}c and \ref{fig5}d). The possibility of tuning the sensorial delay can be exploited, for example, in a search-and-rescue task by setting initially a negative delay so that the robots can thoroughly explore the environment and, at a later stage, a positive delay so that the robots can be collected into clusters to share the gathered information. Collecting all robots can also be easily achieved by sending a strong signal capable of eclipsing the signals emitted by the robots themselves. 

It is also possible to adjust the behavior of the agents by altering the intensity-speed relation to something different than Eq.~(\ref{eq:v(I)}). For example, instead of a monotonically decreasing relation, it is possible to use a relation with a minimum at some specific value. As can be seen in Fig.~S2  in the Supplementary Information, this alters the agent's behavior so that it spends more time where the intensity corresponds to the minimum speed. In this way, it is possible to control where the agent will spend most of its time, which may be useful, e.g., for targeted delivery. Furthermore, in the presence of multiple agents capable of emitting a radially decaying intensity field, changing the intensity-speed relation permits one to control various features of the clusters such as their characteristic size, as shown in Fig.~S3 in the Supplementary Information.

Our results can also be extended to the three-dimensional case, where they still hold with only minor adjustments. This could be important when considering airborne objects (e.g. drones, flying insects, birds) or underwater objects (e.g. fish, submarine robots). In three dimensions, the autonomous agent motion can be modelled by the set of equations
\begin{equation}\label{eq:3d}
\left\{
\begin{array}{ccl}
\displaystyle \frac{dx_t}{dt} 
&=&  
\displaystyle v\left( I_{t-\delta} \right) \sin\theta_t \cos\phi_t \\[12pt]
\displaystyle \frac{dy_t}{dt} 
&=& 
\displaystyle v\left( I_{t-\delta} \right) \sin\theta_t \sin\phi_t \\[12pt]
\displaystyle \frac{dz_t}{dt} 
&=& 
\displaystyle v\left( I_{t-\delta} \right) \cos\theta_t \\[12pt]
\displaystyle \frac{d\theta_t}{dt} 
&=&
\displaystyle \frac{1}{\tau} \cot\theta_t + \sqrt{\frac{2}{\tau}} \, \eta^{(1)}_t \\[12pt]
\displaystyle \frac{d\phi_t}{dt} 
&=&
\displaystyle \frac{1}{\sin\theta_t} \sqrt{\frac{2}{\tau}} \, \eta^{(2)} _t
\end{array}
\right.
\end{equation}  
where $(x_t,y_t,z_t)$ is the position of the agent at time $t$, $\theta_t$ and $\phi_t$ are its azimuthal and polar orientations respectively, and $\eta^{(1)}_t$ and $\eta^{(2)}_t$ are independent white noises. \gv{Similar equations but without delay have already been considered, e.g. in Ref.~\citenum{grossmann2015geometric} to describe active Brownian motion in three dimensions.} The last two equations describe (accelerated) Brownian motion on the surface of the unit sphere (see Supplementary Information). From this model we obtain the approximate steady-state probability distribution (see Appendix~\ref{app:mathematicalderivation} and Supplementary Information), which exists and equals
\begin{equation}\label{eq:density:3d}
\rho_0(x,y,z) = \frac{1}{M \; v(x,y,z)^{1+2\frac{\delta}{\tau}}} \; ,
\end{equation} 
provided that the normalization constant 
$$M = \int v(x,y,z)^{-\left(1+ 2\frac{\delta}{\tau}\right)}\,dx\,dy\,dz < \infty.$$ 
Comparing Eq.~(\ref{eq:density:3d}) and Eq.~(\ref{eq:density}), we note that the main difference is that in the three-dimensional case the uniform distribution occurs for $\delta = -0.5 \tau$ instead of for $\delta = -\tau$. Otherwise, the agents still exhibit a qualitatively different behavior for positive and negative sensorial delay, corresponding, respectively, to an effective drift towards high-intensity and low-intensity regions, as illustrated in Figs.~S4 and S5 in the Supplementary Information. As in the two-dimensional case, also in the three-dimensional case it is possible to engineer this drift by changing the time delay in order to tune the collective behavior of a swarm from aggregation and clustering to segregation.

\section{Conclusion}

We have demonstrated the use of delayed sensorial feedback to control the organization of an ensemble of autonomous agents. We realized this model experimentally by using autonomous robots, further backed it up with simulations, and finally provided a mathematical analysis which agrees with the results obtained in the experiments and simulations. Our findings show that a single robot, measuring the intensity locally, spends more time in either a high or a low-intensity region depending on its sensorial delay. Tuning the value of the delay permits one to engineer the behavior of an ensemble of robots so that they come together or separate from each other.  The robustness and flexibility of these behaviors are very promising for applications in the field of swarm robotics \cite{Bonabeau2000Nature,Dorigo2004AR,Palacci2013Science,Rubenstein2014Science,Werfel2014Science} as well as in the assembly of nanorobots, e.g., for targeted delivery within tissues. Furthermore, since some living entities, such as bacteria, are known to respond to temporal evolution of stimuli \cite{Berg2004Book,Segall1986PNAS}, the presence of a sensorial delay could also explain the swarming behavior of groups of living organisms.

\begin{acknowledgments}
The authors thank Gilles Caprari (GCtronic) for his help with the robots. GV thanks Holger Stark for useful discussions that led to the original idea for this work. JW and AM were partially supported by the NSF grants DMS 1009508 and DMS 0623941. GV was partially supported by Marie Curie Career Integration Grant (MC-CIG) PCIG11 GA-2012-321726 and a Distinguished Young Scientist award of the Turkish Academy of Sciences (T\"UBA). MM was partially supported by T\"UBITAK grant 113Z556.
\end{acknowledgments}

\appendix

\section{Mathematical derivation}\label{app:mathematicalderivation}
We studied the limit of the system (\ref{eq:system:delay}) as $\delta$, $\tau \rightarrow$ 0 at the same rate so that $\delta = c \epsilon$ and $\tau = k \epsilon$ where $c$ and $k$ remain constant in the limit $\delta, \tau, \epsilon \rightarrow 0$.  We expanded $v$ about $t$ to first order in $\delta$ and solved the resulting equations for $\dot{x}$ and $\dot{y}$.  We expanded the resulting system to first order in the small parameter $\frac{\delta}{\sqrt{\tau}}$.  We then considered the corresponding backward Kolmogorov equation for the probability density $\rho$.  We expanded $\rho$ in powers of the parameter $\sqrt{\epsilon}$, i.e. $\rho = \rho _0 + \sqrt{\epsilon} \rho _1 + \epsilon \rho _2 + ...$, and used the standard multiscale expansion method \cite{Pavliotis2008Book} to derive the backward Kolmogorov equation for the limiting density $\rho _0$:
\begin{equation}\label{methods:limitingSDE2d}
{\partial \rho_0 \over \partial t} = \frac{\tau}{2} \left( 1 - \frac{\delta}{\tau} \right) v \left( \frac{\partial v}{\partial x} {\partial \rho_0 \over \partial x} + \frac{\partial v}{\partial y} {\partial \rho_0 \over \partial y} \right) + \frac{\tau v^2}{2} \Delta \rho_0 \; .
\end{equation}
From this equation, we got the limiting SDE:
\begin{equation}\label{eq:djodsms}
\left\{
\begin{array}{ccl}
dx_t = \frac{\tau}{2} \left( 1 - \frac{\delta}{\tau} \right) v (x_t, y_t)  \frac{\partial v}{\partial x} (x_t, y_t) dt + \sqrt{\tau} v (x_t, y_t) dW^1 _t \\
dy_t = \frac{\tau}{2} \left( 1 - \frac{\delta}{\tau} \right) v (x_t, y_t)  \frac{\partial v}{\partial y} (x_t, y_t) dt + \sqrt{\tau} v (x_t, y_t) dW^2 _t
\end{array}
\right.
\end{equation}
where $W^1$ and $W^2$ are independent Wiener processes. Assuming that $v$ is rotation-invariant, we got from Eq.~(\ref{eq:djodsms}) the formula for the radial drift [Eq.~(\ref{eq:drift})]:
\begin{equation}
D(r) = \frac{\tau}{2} \left( 1 - \frac{\delta}{\tau} \right) v(r) \frac{d v}{d r} (r) + \frac{\tau v(r)^2}{ r} \; .
\end{equation}
Setting the right-hand side of the forward (Fokker-Planck) equation corresponding to Eq.~(\ref{methods:limitingSDE2d}) equal to zero, we got the formula for the stationary probability density $\rho _0$ (if it exists) [Eq.~(\ref{eq:density})]
\begin{equation} 
\rho _0 (x, y) = \frac{1}{N \; v(x, y)^{1 + \frac{\delta}{\tau}}} \; ,
\end{equation}
where $N$ is the normalization constant.
A similar analysis follows for the three-dimensional case, leading to the three-dimensional stationary probability density given by Eq.~(\ref{eq:density:3d}).
A more detailed derivation is provided in the Supplementary Information.

\end{document}